\documentclass[10pt,letterpaper]{article}
\usepackage[latin9]{inputenc}
\usepackage{color}
\usepackage{amsmath}
\usepackage{amssymb}
\usepackage{graphicx}
\usepackage[bookmarks=false,
 breaklinks=false,pdfborder={0 0 1},backref=false,colorlinks=true]
 {hyperref}
\hypersetup{
 citecolor=blue,urlcolor=blue}

\makeatletter

\pdfpageheight\paperheight
\pdfpagewidth\paperwidth

\usepackage{icton}
\usepackage{capt-of}
\usepackage{booktabs}
\usepackage{microtype}

\usepackage{cite}
\usepackage[printonlyused]{acronym}
\acrodef{DM}{distribution matcher}
\acrodef{Hi-DM}{hierarchical DM}
\acrodef{PS}{probabilistic shaping}
\acrodef{PAS}{probabilistic amplitude shaping}
\acrodef{AWGN}{additive white Gaussian noise}
\acrodef{QAM}{quadrature amplitude modulation}
\acrodef{FEC}{forward error correction}
\acrodef{PAM}{pulse amplitude modulation}
\acrodef{LUT}{look up table}
\acrodef{MB}{Maxwell-Boltzmann}
\acrodef{ESS}{enumerative sphere shaping}
\acrodef{CCDM}{constant composition distribution matching}
\acrodef{IR}{information rate}
\acrodef{SNR}{signal to noise ratio}
\acrodef{PPM}{pulse position modulation}
\acrodef{invDM}{inverse DM}
\acrodef{TX}{transmitter}
\acrodef{RX}{receiver}
\acrodef{BER}{bit error rate}
\acrodef{SER}{symbol error rate}

\makeatother

\begin{document}
\title{Fiber Nonlinearity Mitigation in Coherent Optical Systems}
\author{Stella Civelli$^{\text{1,2,*}}$, \textit{Member, IEEE}, Dario Cellini$^{\text{2}}$,
\textit{Graduate Student Member, IEEE}, Enrico Forestieri$^{\text{2,3}}$,
\textit{Senior Member, IEEE}, Marco Secondini$^{\text{2,3}}$, \textit{Senior Member, IEEE}}
\address{\textsuperscript{1}CNR-IEIIT, Via Caruso 16, 56122, Pisa, Italy,
\emph{stella.civelli@cnr.it}\\
\textsuperscript{2} TeCIP Institute, Scuola Superiore Sant'Anna,
Via G. Moruzzi 1, 56124, Pisa, Italy\\
\textsuperscript{3}PNTLab, CNIT, Via G. Moruzzi 1, 56124, Pisa, Italy}
\maketitle
\begin{abstract}
Fiber nonlinearity represents a critical challenge to the capacity
enhancement of modern optical communication systems. In recent years,
significant research efforts have focused on mitigating its impact
through two complementary approaches. On the one hand, researchers
have investigated practical digital signal processing (DSP) techniques
to mitigate or compensate for nonlinear impairments, such as reversing
fiber propagation effects through digital backpropagation (DBP). However,
the high computational complexity of these techniques often discourages
their practical implementation. On the other hand, information-theoretic
studies have sought to establish the capacity limits of the nonlinear
optical fiber channel, providing a framework for evaluating the ultimate
performance of existing optical networks and guiding the design of
next-generation systems. This work reviews recent advances and proposes
future directions for nonlinearity compensation and mitigation, including
constellation shaping techniques and low-complexity DBP. Furthermore,
it highlights the potential of these innovations both in advancing
the theoretical understanding of fiber capacity limits and in enabling
practical DSP implementations.
\end{abstract}


\section{INTRODUCTION}

\vspace{-1mm}

Optical fiber carries most of global data traffic, and network capabilities
must keep pace with the growing demand for connectivity. While substantial
research focuses on increasing bit rates through spatial multiplexing
or technological advances---such as special fibers---maximizing
the performance of existing networks also deserves significant attention
\cite{agrell2024roadmap}. In this context, simply increasing signal
power to improve the received signal-to-noise ratio (SNR) is ineffective,
as fiber Kerr nonlinearity degrades the performance of conventional
nonlinearity-unaware systems. Furthermore, Kerr nonlinearity interacts
with  linear effects in a complex way, resulting in a propagation
model that is challenging to handle. Consequently,  recent years have
seen extensive research into boththe fundamental limits of optical
networks and  practical techniques to mitigate fiber nonlinearity
\cite{Essiambre:JLT0210,dar_JLT2017_nonlinear,secondini2019JLT}.

This work focuses on two complementary approaches to nonlinearity
mitigation in coherent optical systems. Digital backpropagation (DBP),
which inverts propagation effects digitally, has been studied for
many years and provides insight into the potential of intra- and inter-channel
nonlinearity compensation. In contrast, constellation shaping is a
more recent technique that aims to encode  information on signals
that are inherently more resilient to fiber nonlinearity. 

\section{DIGITAL BACKPROPAGATION}

\vspace{-1mm}

DBP compensates for linear and nonlinear effects by virtually propagating
the signal backward through the fiber. It can be applied at the transmitter
(TX), receiver (RX), or split between the two. Among the various DBP
techniques,  the most popular is the split-step Fourier method (SSFM)
\cite{agrawal2013_nonlinear_fiber_optics_5e}.

The SSFM emulates backpropagation by dividing the link into many short
steps, assuming linear and nonlinear effects occur independently within
each. Each step consists of a linear operation---implemented as a
frequency-domain filter---and a nonlinear operation---a signal-dependent
phase rotation applied in the time domain.  This  requires  many FFT
and IFFT operations. With sufficient  steps and bandwidth, and complete
knowledge of the signal, SSFM can ideally  undo fiber propagation
effects and recover the transmitted signal. However, there are practical
limitations. If only part of the signal spectrum is available---for
instance, in single-channel DBP for wavelength-division multiplexing
(WDM) systems---interactions with other spectral components cannot
be captured. As a result, only intra-channel nonlinearity is compensated,
while  inter-channel effects must be addressed with complementary
approaches, such as carrier phase recovery or tailored modulation
formats. Additionally, the processing bandwidth and number of steps
affect both accuracy and complexity: higher accuracy requires more
steps and wider processing bandwidth, leading to  increased computation
complexity. For this reason, DBP is generally considered computationally
expensive and impractical.

To address this issue, several methods have been proposed to reduce
DBP complexity while maintaining a good performance. The enhanced
SSFM (ESSFM) uses the same overall structure  but improves the nonlinear
step to account for dispersion--nonlinearity interaction, making
the phase rotation depend not only on  instantaneous power, but also
on neighboring symbols. This improves the accuracy of each step, reducing
the number of required steps  and, thus, the computational cost \cite{Sec:ECOC14,secondini_PNET2016}.
Building on the ESSFM, the coupled-channel ESSFM \cite{civelli2021coupled,civelli2021ISWCS2021}
and the coupled-band ESSFM (CB-ESSFM) \cite{civelli2024CBESSFMECOC,civelli2025new}
were respectively proposed to account for the impact of adjacent channels
or to use subband processing. The CB-ESSFM also includes an optimized
splitting of linear and nonlinear effects in each step, further increasing
the accuracy and  yielding significant gain with feasible computational
complexity.

Recently, machine learning (ML) has been explored for DBP, exploiting
its strength  as a universal function approximator to handle problems
with no closed-form solutions, such as  the nonlinear Schr�dinger
equation. Remarkably,  the SSFM structure closely resembles a deep
neural network (NN), with alternating linear operations and nonlinear
activation functions \cite{hager2020physics}. Learned DBP (LDBP)
uses a large number of low-complexity linear steps---finite impulse
response filters with few coefficients optimized with ML tools---and
models the Kerr effect in the nonlinear activation function \cite{hager_OFC_2018learnedDBP}.
LDBP achieves high accuracy with remarkably short filters  and low
complexity, and was later improved with subband processing \cite{hager2018wideband}.
ML has also been proposed for DBP using learned Volterra-series-based
models, which represent nonlinear effects additively and  enable low-latency
implementations,  enhancing the efficiency of multiple-input  multiple-output
equalization \cite{Castro2025}. 

\section{CONSTELLATION SHAPING}

\vspace{-1mm}

Constellation shaping maps information bits onto constellations of
symbols with optimized geometry and/or probability, aiming to enhance
 overall system performance. Probabilistic amplitude shaping (PAS)
induces a desired probability distribution on the symbol amplitudes
by using a distribution matcher (DM), integrates forward error correction
(FEC), and provides fine rate granularity \cite{bocherer2015bandwidth}.
The DM should be designed to minimize  complexity while accurately
approximating the desired distribution---typically the Maxwell--Boltzmann
(MB) distribution for AWGN channels. Both the constant composition
DM (CCDM) \cite{schulte2016CCDM} and the hierarchical DM (HiDM) \cite{yoshida2019hierarchicalDM,civelli2020entropy}
can provide arbitrary amplitude distributions, with the latter offering
an excellent trade-off between accuracy and complexity. Sphere shaping
(SpSh), implementable via  enumerative SpSh, minimizes the mean energy
per symbol for a given rate, constellation order, and block length,
hence achieving  the best performance over  the AWGN channel with
an average-power constraint \cite{gultekin2018Sphereshaping,gultekin2020probabilistic}.

Recent research has explored the use of symbol sequences that are
resilient to fiber nonlinearity to improve system performance.  Short-block-length
DMs  yield a nonlinear shaping gain due to reduced energy fluctuations
among amplitudes belonging to the same DM block. The optimal block
length---and thus the balance between linear and nonlinear shaping
gain---depends on the DM design: longer blocks improve linear shaping
gain, while shorter blocks favor nonlinear shaping gain.  Among conventional
PAS strategies, SpSh provides the largest shaping gain in most scenarios
\cite{civelli2023JLTNPN}, and modified versions of SpSh have been
designed to optimize the nonlinear shaping gain in some specific scenarios
\cite{gultekin2021kurtosis,gultekin2022band}.However, the gain provided
by short-block-length DMs is largely lost when a carrier phase recovery
(CPR) is included in the system \cite{civelli2023JLTNPN}. Indeed,
CPR compensates for slow phase variations caused not only by laser
phase noise but also by inter-channel nonlinearities---effects that
are likewise mitigated by reducing signal intensity fluctuations.

The limited gains achieved with conventional PAS have motivated researchers
to extend constellation shaping to multi-dimensional shaping with
sequence selection, which improves performance by transmitting  \emph{good}
symbol sequences, properly selected to minimize the impact of fiber
nonlinearity  in the desired scenario \cite{civelli2021sequenceECOC,civelli2024sequence}.
Ideal sequence selection was used to establish  a new capacity lower
bound for dual-polarization (DP) WDM systems and to assess the ultimate
potential of nonlinear shaping \cite{secondini2022new}. Several practical
sequence selection methods have since been proposed, including bit
scrambling (BS) sequence selection \cite{civelli2023practicalOFC,civelli2024sequence},
the  method proposed in \cite{askari2023probabilistic}, and list-encoding
CCDM \cite{wu2022list}. In the general approach, a sequence of information
bits is mapped to $N_{t}$ candidate sequences of $n$ 4D-symbols;
the sequences are compared by using a proper metric;  and the \emph{best}
one is selected for transmission. This results in an acceptance rate
of $1/N_{t}$ and a rate loss  of $\log_{2}(N_{t})/n$ bits/4D. The
combination of FEC and sequence selection has been investigated in
\cite{civelli2023practicalOFC}. The key element of sequence selection
is the metric, which should be accurate enough to  assess the quality
of the sequences, and simple enough to allow practical implementations,
as it should be evaluated $N_{t}$ times. Remarkably, sequence selection
outperforms PAS based on SpSh with optimized blocklength only if the
metric accounts for the signs of the sequences---not just their amplitudes
(energy). A few metrics that accounts for  both amplitudes and signs
have been proposed, including the dispersion-aware energy dispersion
index \cite{liu2024sequence}, the lowpass-filtered symbol-amplitude
sequence \cite{askari2023probabilistic}, and the nonlinear interference
metric. The latter can be efficiently implemented using the CB-ESSFM
to simulate channel propagation with a small number of steps. A preliminary
analysis of the trade-off between computational cost and shaping gain
for BS sequence selection with CB-ESSFM showed that modest gains can
be achieved with reasonable complexity, but larger gains require an
impracticable cost, highlighting the need for more efficient metrics
for sequence selection \cite{civelli2024costOFC}.

\section{NONLINEARITY MITIGATION: GAINS AND COMPLEXITY}

\vspace{-1mm}

This manuscript considers the same setup as \cite{civelli2024sequence}.
The TX maps information on DP $64$-QAM symbols, either with with
uniform (U) distribution or PAS (SpSh with blocklength 256) with rate
$9.2$bits/4D, eventually followed by sequence selection. The symbols
modulate root-raised cosine pulses at $46.5$~GBd, forming a DP-WDM
signal with $5$ channels spaced $50$GHz apart. The signal propagates
through a link consisting of $30\times100$km spans of single-mode
fiber, each followed by an erbium-doped fiber amplifier with noise
figure of $5$dB, which exactly compensates for the span loss. At
the RX, the central channel is demultiplexed, undergoes dispersion
compensation or DBP, matched filtering, and symbol time sampling.
Eventually,  mean phase removal is applied when the TX and RX laser
linewidth is $\ell_{w}=0$,  while CPR is performed using blind phase
search ($481$ symbols) when $\ell_{w}=100$kHz. The system performance
is measured in terms of spectral efficiency (SE) in bit/s/Hz.

\begin{figure}[h]
\begin{centering}
\vglue-6pt 
\par\end{centering}
\centering{}\includegraphics[width=0.5\textwidth]{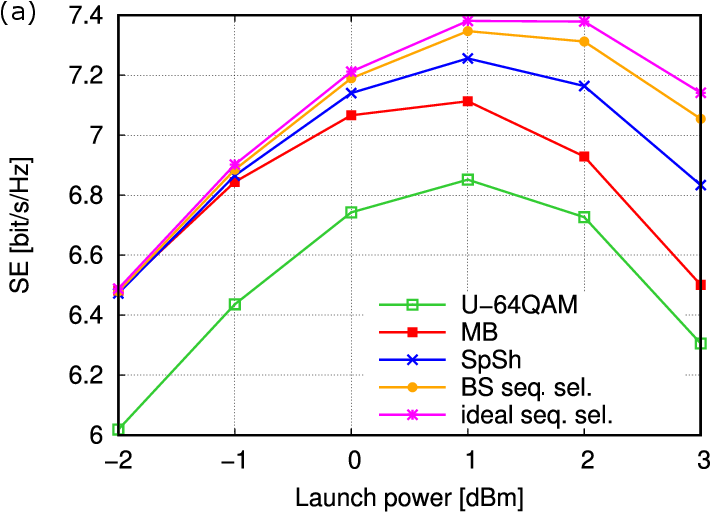}\includegraphics[width=0.5\textwidth]{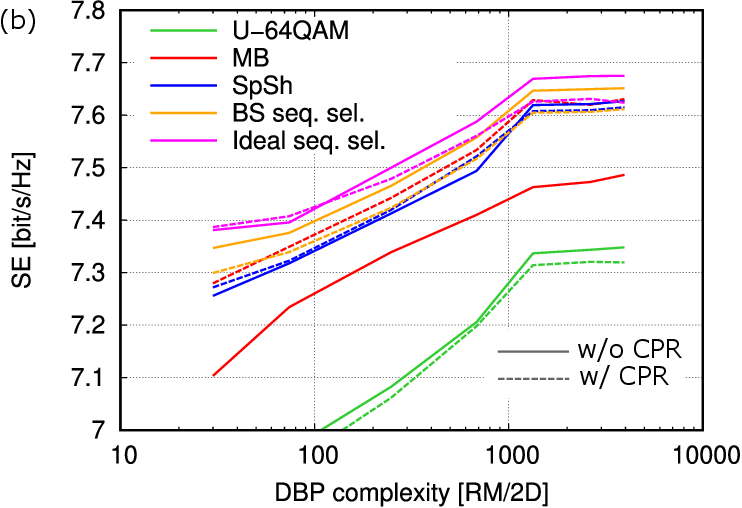}\caption{\label{fig:results}(a) SE versus power for different modulation techniques
(dispersion compensation, without CPR) (b) SE versus DBP complexity,
with and without CPR}
\vglue-6pt \vspace*{-1pc}
\end{figure}

Figure \ref{fig:results}\,(a) shows the SE versus launch power for
$\ell_{w}=0$ and dispersion compensation, comparing different modulation
schemes:  U-64QAM; PAS with i.i.d. MB-distributed symbols (the ideal
target distribution for the AWGN channel); PAS implemented via SpSh
with an optimized blocklength of $256$ amplitudes; and the same SpSh-based
PAS followed by either BS sequence selection ($N_{t}=256$, $n=512$,
NLI metric computed via CB-ESSFM with 1 band, 30 steps, and 1.125
samples/symbol \cite{civelli2025new,civelli2024costOFC}) or  ideal
sequence selection ($N_{t}=256$, $n=512$, NLI metric evaluated with
many SSFM steps accounting for inter-block interference \cite{civelli2024sequence}).
The results show that (i) PAS techniques yield a $\approx0.5$bits/s/Hz
gain  over  U-$64$QAM in the linear regime; (ii) SpSh matches the
 performance of the ideal MB distribution in the linear regime and
outperforms it in the nonlinear regime, with a gain of $0.15$bit/s/Hz;
(iii) sequence selection  further improves  performance by 0.09bit/s/Hz
with the practical implementation, and $0.13$bit/s/Hz with   the
ideal one.

Figure \ref{fig:results}\,(b) shows the optimal performance of the
aforementioned modulation techniques as a function of  DBP complexity
(real multiplications per 2D symbol, RM/2D), where DBP is implemented
using  CB-ESSFM ($1$ band, $1.125$ samples/symbol) with varying
number of steps \cite{civelli2025new}. Results for $\ell_{w}=0$Hz
(no CPR) and $\ell_{w}=100$kHz (with CPR) are shown with solid and
dashed lines, respectively. The figure shows that (i) all modulation
techniques benefit similarly from increased DBP complexity, regardless
of  CPR; (ii) the nonlinear shaping gain from short-block-length SpSh
vanishes with CPR (MB performs as SpSh); (iii) the nonlinear shaping
gain from sequence selection remains even with CPR, highlighting its
potential for nonlinear constellation shaping. This results illustrate
the trade-offs among  different DSP techniques for nonlinearity mitigation.
For example, SpSh  achieves the same performance as U-64QAM with DBP
at $\approx1000$RM/2D ($N_{\text{st}}=30$), while practical sequence
selection provides an additional gain of $\approx0.1$bit/s/Hz over
SpSh at the same DBP complexity. 

\section{CONCLUSIONS AND OUTLOOK}

\vspace{-1mm}

This work assessed the impact of constellation shaping  and DBP on
the performance of coherent optical transmission systems. DBP effectively
compensates for intra-channel nonlinearity, regardless of the employed
modulation and the presence of CPR. Moreover, recent advances have
significantly reduced its complexity, enabling practical implementations.
PAS achieves notable gains in the linear regime and additional gains
in the nonlinear regime; however, these latter gains become irrelevant
when CPR is included. In contrast, sequence selectionenables shaping
that is more precisely tailored to nonlinearity mitigation, achieving
higher gains that persist even in the presence of CPR. While this
approach provides insight into  the potential of nonlinear shaping
and guides the development  of future modulation techniques, its current
complexity remains an obstacle to deployment.

\section*{ACKNOWLEDGMENTS}

\vspace{-1mm}

This work was partially supported by the European Union - Next Generation
EU under the Italian National Recovery and Resilience Plan (NRRP),
Mission 4, Component 2, Investment 1.3, CUP J53C22003120001, CUP B53C22003970001,
partnership on \textquotedblleft Telecommunications of the Future\textquotedblright{}
(PE00000001 - program \textquotedblleft RESTART\textquotedblright )


\end{document}